\documentstyle[preprint,aps,epsf]{revtex}

\begin{document}

\title{Non-resonant direct {\it p}- and {\it d}-wave 
       neutron capture by ${}^{12}$C}

\author{
T.~Kikuchi$^a$, Y.~Nagai$^a$, T.~S.~Suzuki$^a$, 
   T.~Shima$^a$, T.~Kii$^a$, M.~Igashira$^b$, \\
A.~Mengoni$^{c,d,e}$, and T.~Otsuka$^{c,e}$
}

\address{
$^a$ Tokyo Institute of Technology, 
     Department of Applied Physics, \\
     O-okayama, Meguro, Tokyo 152, Japan\\
$^b$ Tokyo Institute of Technology, 
     Research Laboratory for Nuclear Reactors, \\
     O-okayama, Meguro, Tokyo 152, Japan\\
$^c$ The Institute of Physical and Chemical Research (RIKEN), \\
	Wako, Saitama 351-01, Japan \\
$^d$ National Institute for New Technologies, 
     Energy and Environment (ENEA),\\
        v.~Don Fiammelli 2, I-40128 Bologna, Italy \\
$^e$ The University of Tokyo,
     Department of Physics,  \\
     Hongo, Bunkyo, Tokyo 113, Japan\\
}

\maketitle

\begin{abstract}
  
Discrete $\gamma$-rays from the neutron capture state 
of ${}^{13}$C to its low-lying bound states have been measured 
using pulsed neutrons at $E_{n}$ = 550 keV.
The partial capture cross sections have been determined 
to be 1.7$\pm$0.5, 24.2$\pm$1.0, 2.0$\pm$0.4 and 1.0$\pm$0.4 $\mu$b 
for the ground ($J^{\pi}=1/2^{-}$), first ($J^{\pi}=1/2^{+}$), 
second ($J^{\pi}=3/2^{-}$) and third ($J^{\pi}=5/2^{+}$) 
excited states, respectively.  
From a comparison with theoretical predictions based on 
the non-resonant direct radiative capture mechanism,  
we could determine the spectroscopic factor 
for the  $J^{\pi} = 1/2^{+}$ state to be 0.80 $\pm$ 0.04, 
free from neutron-nucleus interaction ambiguities in the continuum. 
In addition we have detected the contribution of the non-resonant 
{\it d}-wave capture component in the partial cross sections 
for transitions leading to the $J^{\pi} = 1/2^{-}$ 
and $J^{\pi} = 3/2^{-}$ states. 
While the {\it s}-wave capture dominates at $E_{n} \le $ 100 keV,
the {\it d}-wave component turns out to be very
important at higher energies. From the present investigation
the ${}^{12}$C$(n,\gamma)^{13}$C reaction rate is obtained
for temperatures in the range $10^{7} - 10^{10} \ {}^\circ$K.
\end{abstract}


\section{Introduction}
An increasing interest has been recently developed around the low energy 
neutron capture reaction mechanisms induced by light nuclei 
in view of the importance of the related cross section
for the investigation of nuclear structure properties
as well as for applications 
in nuclear astrophysics\cite{Nax91,Ohx94,Igx95,Hox91,Otx94,Mex95}. 
In fact, it has been recently shown that $(n,\gamma)$
reactions, together with their time-reversal counterpart,
the Coulomb dissociation process, can provide crucial
information on the structure of bound-state
wave functions in light nuclei. 
In particular, it has been recently shown\cite{Mex95,Mex97}
that the external component of the radial wave function
can be extracted from the matrix elements of the 
electromagnetic E1-operator connecting {\it p}-wave 
states in the continuum to loosely-bound {\it l=0} orbits
(halo states).

On the other hand, neutrons with kinetic energies 
in the keV-range correspond to the temperature in stars 
where a variety of capture processes take place 
(1 keV = 11.6 $\times 10^6 \ {}^{\circ}$K). The 
capture cross sections of medium-mass and heavy nuclei 
required for the interpretation of various mechanisms 
of stellar nucleosynthesis 
have been extensively measured in the course
of the last few decades \cite{MaGi65,Kax89}.
However, fast neutron capture cross sections have
been rarely measured for light nuclear targets.

Recently, discrete $\gamma$-rays following the neutron capture 
by light nuclei such as ${}^{12}$C and ${}^{16}$O 
have been successfully measured for incident
neutron energies in the keV region \cite{Nax91,Ohx94,Igx95}.  
It was found that the partial cross section 
from the capture state of ${}^{13}$C to the ground state 
($J^{\pi} = 1/2^{-}$) decreases with increasing 
neutron energy as expected for {\it s}-wave capture, 
while the cross sections leading to the  
first-excited states ($J^{\pi} = 1/2^{+}$) 
in ${}^{13}$C and ${}^{17}$O 
increase with increasing neutron energy. In addition,
the capture $\gamma$-ray branching ratios    
markably favor the transition to these excited states, 
unlike the observed branching ratios 
of thermal neutron capture.  
These new observations indicate the characteristic 
features of a non-resonant {\it p}-wave 
direct neutron capture process taking place 
as will be discussed below.  

The discrete $\gamma$-ray detection is crucial 
for studying nuclear structure and reaction mechanisms 
since the $\gamma$-ray carries unique information 
connecting the capture state and the low-lying 
bound states. 
However, the prompt discrete $\gamma$-ray detection 
for capture events in light nuclei is
extremely difficult due to the small cross
section expected for light nuclei.
As an example, extrapolating 
the thermal capture cross section
of ${}^{12}$C, $\sigma_{n,\gamma}^{th}$ = 3.53 mb,
with a $1/v$-law one obtains 
$\sigma_{n,\gamma}$ = 3.2 $\mu$b at $E_{n}$ = 30 keV 
and  $\sigma_{n,\gamma}$ = 0.8 $\mu$b at $E_{n}$ = 500 keV.
In actual experiments it is difficult to discriminate 
the weak $\gamma$-ray signal from the huge background 
due to the capture of neutrons scattered and/or 
thermalized by the collision with various materials 
in the experimental room.
In spite of this, we have recently 
succeeded to detect discrete $\gamma$-rays 
from the ${}^{12}$C$(n,\gamma)^{13}$C reaction
for incident neutron energies 
in the keV region by developing a highly sensitive 
discrete $\gamma$-ray detection system.
We have already reported on the measurement
performed at neutron energies up to $E_{n}$ = 200 keV
\cite{Ohx94}. Here we report on the results obtained
at much higher energy, $E_{n}$ = 550 keV.
From this new experiment and from the theoretical
investigation which we will describe below, we
have derived the ${}^{12}$C$(n,\gamma)^{13}$C reaction rate 
for temperatures in the range $10^{7} - 10^{10} \ {}^{\circ}$K.

\section{The experiment}
\begin{figure}
\begin{center}
\leavevmode
\hbox{
\epsfxsize=13.0cm
\epsfysize=8.7cm
\epsffile{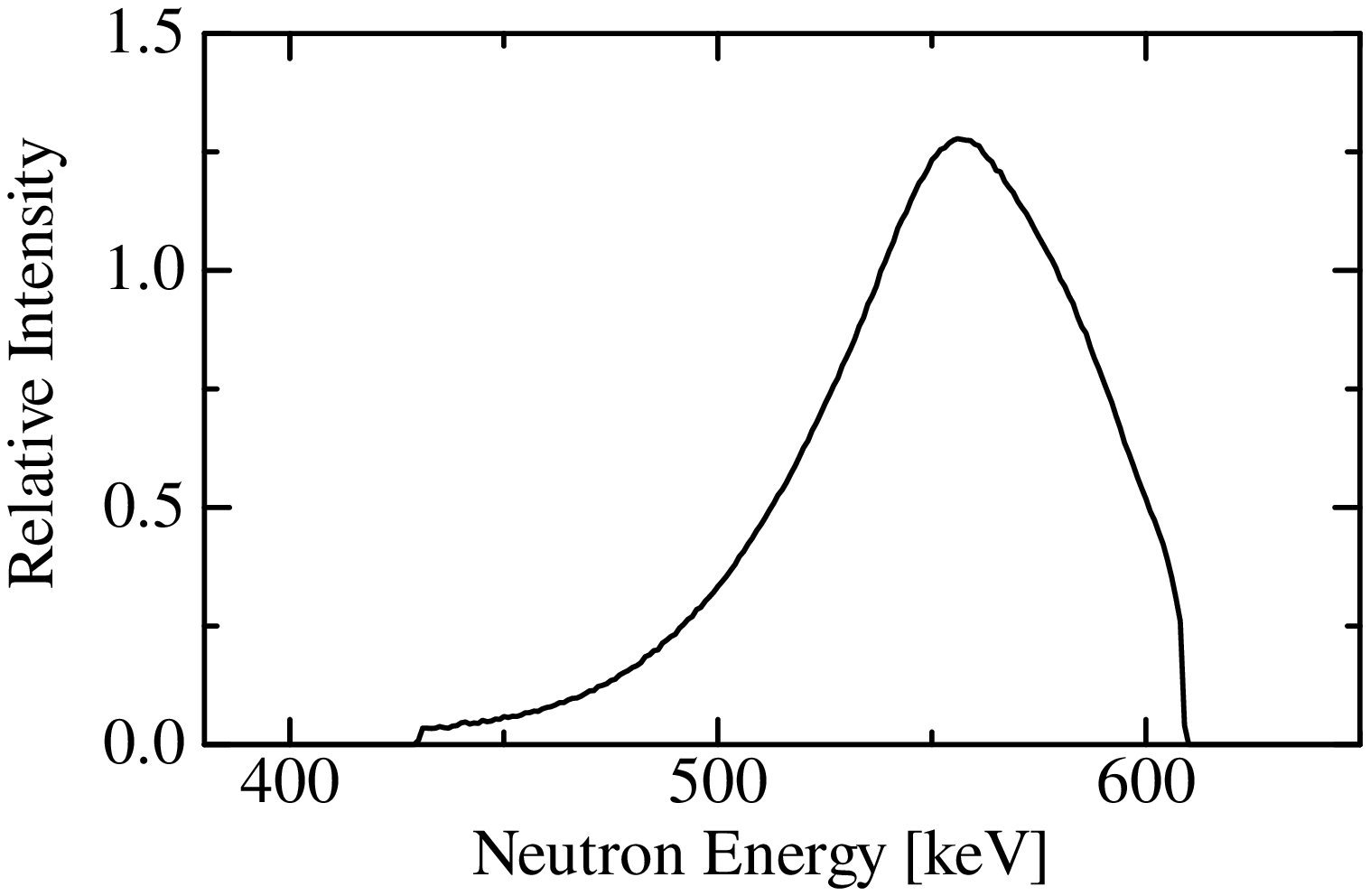}}
\end{center}
\caption{Neutron energy spectrum measured by a
${}^{6}$Li-glass scintillation counter with a TOF method.
The spectrum has been corrected for the energy-dependendent
neutron detection efficiency of the ${}^{6}$Li-glass.}

\label{fig1}
\end{figure}

The experiment was carried out using 
a prompt discrete $\gamma$-ray detection method 
combined with a pulsed neutron beam.  
The neutrons were produced 
by the ${}^{7}$Li$(p,n){}^{7}$Be reaction 
using the pulsed proton beam 
with a pulse width of 1.5 ns, 
provided by the 3.2 MV Pelletron Accelerator 
of the Research Laboratory for Nuclear Reactors 
at the Tokyo Institute of Technology.  
An average beam current of 12 $\mu$A was obtained 
at a repetition rate of 4 MHz.  
The neutron energy spectrum was measured 
by a ${}^{6}$Li-glass scintillation counter
set at 9.6${}^{\circ}$ 
with respect to the proton beam direction.
An averaged neutron energy 
spectrum of 550 keV
(shown in Fig.~\ref{fig1} after correction
for energy dependent efficiency detection) 
was obtained at the sample position as a consequence 
of the reaction kinematics.    
Two samples with a 90 mm diameter, 
one of natural carbon (C) of reactor grade 
and one of gold (Au) were placed 19.8 cm away 
from the Li target at 0${}^{\circ}$ with respect 
to the proton beam direction.  
This distance was necessary to separate 
the $\gamma$-ray events caused by the 
sample from the intense $\gamma$-ray flux coming 
from the ${}^{7}$Li$(p,\gamma){}^{8}$Be 
reaction taking place at the neutron target position.   
The Au sample was used for normalization of the absolute 
capture cross section since the cross section of Au 
is known accurately (within an uncertainty of 3\% \cite{ENDFB}).

The thickness of the C sample was 30 mm. 
This has been determined so as to balance 
the reaction yield with the neutron transmission.  
In fact, with increasing sample thickness the yield increases, 
but the neutron transmission becomes lower.  
Therefore, the number of events due to multiple scatterings 
in the sample increases and the cross section would
not be determined accurately.

The prompt $\gamma$-rays emitted in the transition
from the capture state of ${}^{13}$C 
to the low-lying states were measured by an anti-Compton NaI(Tl) 
spectrometer \cite{Igx94} consisting of
a central NaI(Tl) detector with a diameter 
of 6 inches and length of 8 inches and an annular detector 
with thickness of 3 inches and length of 14 inches.
The spectrometer was set at $125.3^{\circ}$ with
respect to the proton beam direction.  
Since the NaI(Tl) detector is known to be sensitive 
to neutrons, we have heavily shielded it 
using ${}^{6}$LiH, borated paraffin, Cd sheet and Pb.  
Using the specially designed spectrometer 
together with the pulsed neutron beam we could measure 
discrete $\gamma$-rays emitted from a neutron 
capture state with cross sections as small as 
$\approx$ 10 $\mu$b at 30 keV.   
The use of a pulsed neutron beam was crucial 
not only to determine the neutron energies 
(by TOF method) but also to discriminate 
the capture events due to the high energy 
neutrons from those arising from 
scattered and/or thermalized ones. 

It should be noted that when the neutron energy 
is as low as 30 keV, the neutrons produced 
at the neutron target position 
are emitted within a narrow cone 
with respect to the proton beam direction.  
However, when the energy is of the order 
of 500 keV, the cone aperture becomes large and 
the neutrons can be scattered and/or directly captured 
by various materials near the NaI(Tl) detector,
producing background. Since there were materials 
containing carbon atoms in the shield of the NaI(Tl) 
spectrometer (for example borated paraffin), 
we made special efforts to reduce these background
events \cite{Nax97}.  

\begin{figure}
\begin{center}
\leavevmode
\hbox{
\epsfxsize=13.0cm
\epsfysize=8.7cm
\epsffile{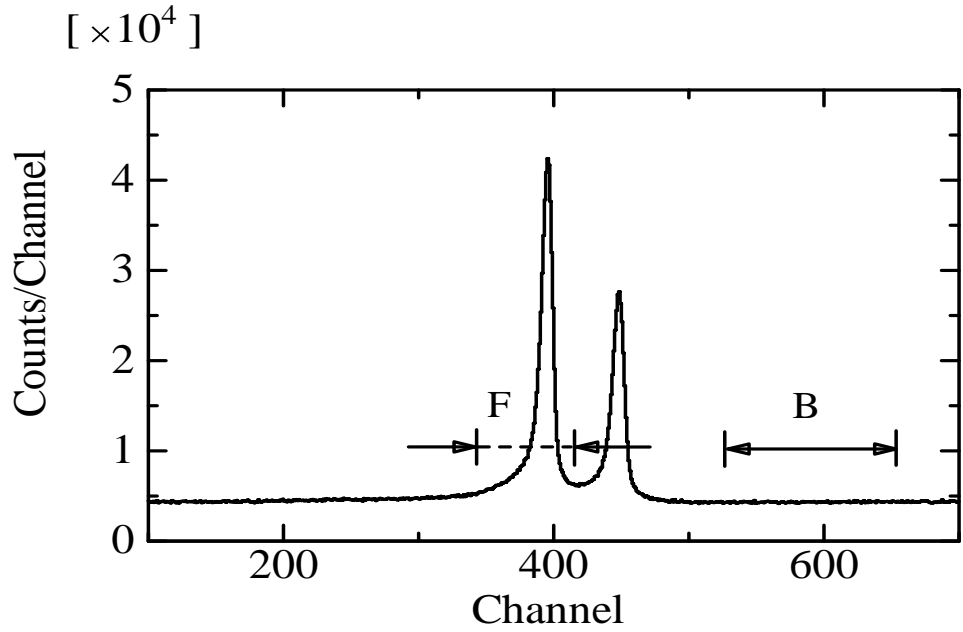}}
\end{center}
\caption{Time-of-flight spectrum measured by the NaI(Tl)
detector for the Au sample. The sharp peak at channel 440 is
due to the $\gamma$-ray from the ${}^{7}$Li$(p,\gamma)^{8}$Be
reaction. The regions F and B were chosen to obtain the foreground 
and background $\gamma$-ray spectra, respectively (see text). }
\label{fig2}
\end{figure}

\begin{figure}
\begin{center}
\leavevmode
\hbox{
\epsfxsize=13.0cm
\epsfysize=8.7cm
\epsffile{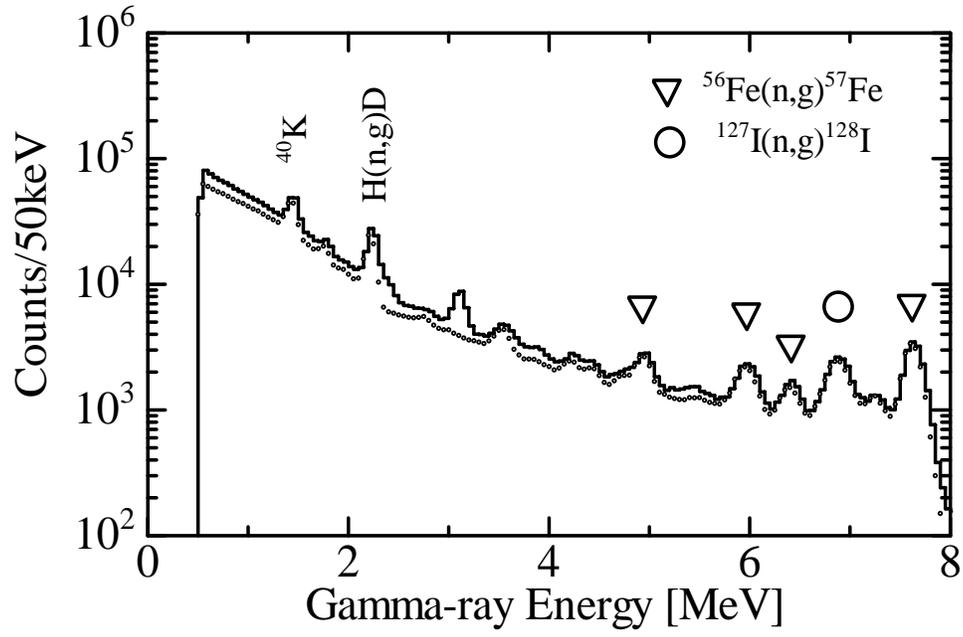}}
\end{center}
\caption{Foreground (solid line) and background (dotted line)
$\gamma$-ray spectra for the  ${}^{12}$C$(n,\gamma)^{13}$C. 
Iron was used to cover the NaI(Tl) spectrometer. }
\label{fig3}
\end{figure}

The TOF spectrum measured by the NaI(Tl) spectrometer 
for the Au sample is shown in Fig.~\ref{fig2}.
The sharp peak at channel 440 
and the broad one at channel 380 are 
due to the ${}^{7}$Li$(p,\gamma){}^{8}$Be  
and to the ${}^{197}$Au$(n,\gamma){}^{198}$Au 
reactions, respectively.  
The foreground (FG) and background (BG) 
spectra for the ${}^{12}$C$(n,\gamma){}^{13}$C
reaction are shown in Fig.~\ref{fig3}.  
They have been obtained by 
putting the gates on F- and B-regions 
in the TOF spectrum shown in Fig.~\ref{fig2}.  
The background subtracted (BS) spectrum is
shown in Fig.~\ref{fig4}. 

The $\gamma$-ray energy calibration curve was made 
using the 1.461 and 2.615 MeV $\gamma$-rays from 
the decays of natural radioactivities of
${}^{40}$K and ${}^{208}$Tl, respectively. 
In addition, several $\gamma$-rays from the 
${}^{56}$Fe$(n,\gamma){}^{57}$Fe reaction
(iron was used to cover the NaI(Tl) spectrometer)
induced by thermal neutrons have been used 
for this purpose \cite{Bix73}.
These neutrons were produced by the collision
of the incident neutron beam with various materials 
in the experimental room.

The continuum background in the BS spectrum was considered 
to be mainly due to the high energy (up to 17 MeV) 
$\gamma$-rays arising from 
the ${}^{7}$Li$(p,\gamma){}^{8}$Be reaction 
at the Li target position. 
In fact, this background level increased as the gate position 
in the TOF spectrum was shifted closer to the peak 
of  the ${}^{7}$Li$(p,\gamma){}^{8}$Be reaction. 
Hence we measured a $\gamma$-ray spectrum without the C sample 
in order to subtract the continuum background.

The net spectrum, free from the continuum part,
is shown in Fig.~\ref{fig5}.  
Two intense peaks at 2.37 and 3.09 MeV can be seen.
They arise from the $\gamma$-ray cascade originating 
from the capture state of ${}^{13}$C leading
to the ground state via the excited state at 3.09 MeV 
($J^{\pi}=1/2^{+}$).  
The peak width of the 2.37 MeV $\gamma$-ray is 
larger than that of the 3.09 MeV
because of the additional contribution 
of the energy spread of the incident neutrons
in the latter transition. 
The small peaks at 5.46, 3.85 and 3.68 MeV are due to
the $\gamma$-rays originating from the capture state,
the third- and the second-excited states respectively
decaying to the ground state (see a partial level scheme 
of ${}^{13}$C in Fig.~\ref{fig6}).  
It should be mentioned here that the latter two 
$\gamma$-rays were not observed clearly in the previous 
experiment at $E_{n} \leq $ 200 keV \cite{Ohx94}.

The intensities of these $\gamma$-rays were analyzed 
by the stripping method using a response function 
obtained experimentally\cite{Igx94}.   
The $\gamma$-ray relative intensities thus obtained 
are compared with those generated by thermal neutrons 
in Fig.~\ref{fig6}.   
It is worthwhile to note here that the $\gamma$-ray strength 
from the capture state feeding directly the first excited state 
in ${}^{13}$C is large for 550 keV neutrons, 
whereas for thermal neutrons the largest strength is
observed for the transition leading directly 
to the ground state\cite{Ajz91}.
The $\gamma$-ray for the transition leading 
to the third excited state was identified for the first time 
for incident neutron energies of 550 keV
by the 3.85 MeV $\gamma$-ray, as discussed above.  

Since the thermal capture proceeds only
by {\it s}-wave neutrons, the strongest transition 
from the capture state to the ground state 
is an E1 transition. 
Assuming the same E1 character for the transition 
leading to the first excited state, 
the capture process must proceed 
via {\it p}-wave neutron capture.  
Among other things it follows that,
because of the E1 character and because 
the measurement is performed at 125.3${}^{\circ}$,
where the second Legendre polynomial is zero, 
the observed intensity represents the angle-integrated 
value.
  
The partial capture cross section $\sigma_{f}$ 
for the ${}^{12}$C$(n,\gamma)^{13}$C reaction 
feeding a low-lying state ({\it f}) is given by using 
the $\gamma$-ray yield $Y_{\gamma,f}$ as 

\begin{equation}
\sigma_{f} = F \ \frac{ (\phi)_{Au} }{ (\phi)_{C} } 
\frac{ (r^{2} n)_{Au} }{ (r^{2} n)_{C} }
\frac{ Y_{\gamma,f}(C) }{  Y_{\gamma,f}(Au) } 
\sigma(Au)
\label{eq1a}
\end{equation}
with
\begin{equation}
F = \frac{ (C_{nm} C_{ns} C_{\gamma a} C_{fs} )_{Au} }
         { (C_{nm} C_{ns} C_{\gamma a} C_{fs} )_{C} }.
\label{eq1b}
\end{equation}

Here, $F$ is the correction factor defined in Eq.~\ref{eq1b} 
and $\phi$ is the neutron yield, respectively.  
$r$ and $n$ are the radius (in cm) 
and thickness (atoms/barn) of the sample.  
$Y_{\gamma}(Au)$ and $\sigma(Au)$
are the $\gamma$-ray yield and 
the absolute capture cross section of Au, respectively.  
The correction factors for neutron multiple scattering effects, 
for the shielding of the incident neutrons in the sample, 
for the $\gamma$-ray absorbed by the sample
and for the finite size of the sample are
indicated by $C_{nm}$, $C_{ns}$, $C_{\gamma a}$ and $C_{fs}$
respectively.  
The factors $C_{nm}$ and $C_{ns}$ were calculated 
by a Monte-Carlo code, TIME-MULTI \cite{Sex94} and
$C_{\gamma a}$ and $C_{fs}$ were also estimated using 
a Monte-Carlo method. These correction factors
are given in Table 1. In calculating the correction factors, 
both the elastic and the capture reaction processes 
have been considered as main sources of neutron attenuation. 
Hence, the coefficient $C_{nm}$ may differ 
for $s$- and for $p$-wave neutrons \cite{Sex94}.

\section{Results and comparison with DRC model calculation}
The final results for the partial capture cross sections 
for the transitions leading to the ground, first, 
second and third excited states 
in the ${}^{12}$C$(n,\gamma)^{13}$C reaction
are given in Table 2.
In Fig.~\ref{fig7} we show the present results together 
with the results of the previous measurement 
performed at incident neutron energies comprised 
between 10 and 200 keV. 
A comparison is made with calculations
based on the non-resonant {\it s}-, {\it p}-
and {\it d}-wave DRC \cite{Mex95} whose description
we will briefly summarize here.

\subsection{DRC model calculation}
It has been recently shown that 
while the DRC process of incident {\it s}-wave neutrons 
captured into bound {\it l=}1 orbits is very sensitive 
to the neutron-nucleus interaction in the continuum, 
the matrix elements for E1 transitions (hence the
cross section) for incident {\it p}-wave neutron capture,
leading to the bound {\it l=}0 or {\it l=}2 orbits 
are insensitive to this interaction \cite{Mex95}.
This can be intuitively understood by simply reminding that 
the {\it p}-wave neutron collision at low energies is essentially 
peripheral and therefore unaffected by the neutron-nucleus 
potential. Consequently, since the basic information 
on the {\it p}-wave neutron capture process is carried 
by the structure of the bound-state, 
its spectroscopic factor, $S_{b}$, can be derived 
from the measured capture cross section.   

To show this let us remind that, for an E1 transition, 
the DRC cross section of incident neutrons with energy $E_{n}$ 
is given by

\begin{equation}
\sigma_{n,\gamma} = \frac{16 \pi}{9 \hbar} k_{\gamma}^{3} \bar{e}^2
\vert Q^{(E1)}_{c \rightarrow b} \vert^{2}
\label{eq2}
\end{equation}

Here, $k_{\gamma} = \epsilon_{\gamma}/ \hbar c$ is the
$\gamma$-ray wave number for a transition with 
emission energy $ \epsilon_{\gamma}$ and $\bar{e} = -Z/A$ 
is the E1 effective charge for a neutron. 
All the quantum numbers necessary to uniquely define 
the initial and the final states have been lumped 
into the notation {\it c} and {\it b}, respectively.  
The matrix elements for the dipole operator 
$\hat{T}^{E1}$ can be decomposed into
the products of three terms

\begin{equation}
Q^{(E1)}_{c \rightarrow b} = < \Psi _{b}
\vert \hat{T}^{E1} \vert \Psi _{c} >
\equiv \sqrt{S_b} \ {\cal I}_{c,b} \ A_{c,b} .
\label{eq3}
\end{equation}

We have indicated with $A_{c,b}$
the angular coupling part (including orbital and
spin coupling coefficients \cite{Mex95}) and 
with ${\cal I}_{c,b}$ the radial part 
of the transition matrix elements.  
The entrance channel wave function 
(the spin component is omitted in the following)

\begin{equation}
\Psi _{l m}(\mbox{\bf r}) \equiv
w_{l}(r)
\frac{Y_{l,m}(\theta,\phi)}{r v^{1/2}} =
\frac{i \sqrt{\pi}}{k} \sqrt{2l+1} i^{l}[I_l - U_{l} O_l]
\frac{Y_{l,m}(\theta,\phi)}{r v^{1/2}}
\label{eq4}
\end{equation}

where 

\begin{equation}
I_{l} \sim \mbox{exp}( -ikr + \frac{1}{2}il\pi )
\quad \mbox{and} \quad
O_{l} \sim \mbox{exp}( +ikr - \frac{1}{2}il\pi )
\label{eq5}
\end{equation}

may be simply replaced by the {\it l=}1 component 
of the partial wave decomposition of a plane-wave 
in the incident channel
 
\begin{equation}
w_{l_{c}=1}(r) = \frac{(2i) \sqrt{3 \pi}}{k} kr j_{1}(kr).
\label{eq6}
\end{equation}

Here, $j_{1}(kr)$ is a spherical Bessel function, 
$k = \sqrt{2\mu E_{n}} / \hbar c$   
is the wave number corresponding to the incoming neutron energy 
$E_{n}$ and reduced mass $\mu$. This ansatz is
equivalent to assuming $U_{l} \equiv 1$ for all $l$.
In the more general situation in which 
the neutron-nucleus interaction in the continuum 
appreciably perturbs the scattering channel 
wave function, the collision matrix
$U_{l}$ can be evaluated numerically using some
(optical) model potential. The scattering matrix
is related to the optical model phase-shift
simply by $U_{l} = \exp{(2i\delta_{l}^{opt.})}$.

The radial part of the overlap integral 

\begin{equation}
{\cal I}_{l_{c}l_{b}} \equiv \int\limits_{0}^{\infty}
u_{l_b}(r) r w_{l_c}(r) dr =
i \sqrt{12 \pi} \int\limits_{0}^{\infty}
j_{1}(kr)  r^{2} u_{l_b}(r) dr
\label{eq7}
\end{equation}

can be easily evaluated numerically 
for any given single-particle radial wave function 
of the final state, leaving the factor $S_{b}$   
as the only quantity in Eq.~\ref{eq3}
to be determined by the experiment.  
The accuracy of $S_{b}$ thus determined 
depends on the reliability of the final state 
wave function and on the accuracy 
of the measured cross section.

The standard method to calculate this wave function 
is to use a single-particle model potential, 
for example a Woods-Saxon potential of type

\begin{equation}
V(r) = \frac{-V_{0}}{ 1 + \exp{[ (r - R)/a ]}}
\label{eq8}
\end{equation}

and solve the bound-state problem for given
parameters $R, V_{0},$ and $a$. The depth
of the potential well, $V_{0}$, can be adjusted 
to reproduce the binding energy. 
We have adopted this procedure in
our calculation assuming the following
geometrical parameters for the potential:
$ R = 2.86 $ fm (corresponding to
$r_{0} = 1.25 $ fm in $ R = r_{0} A^{1/3}$)
and $ a = 0.65 $ fm. A spin-orbit coupling
potential with strength $V_{so} = 6.5 $ MeV
was added in the case of {\it p}- and
{\it d}-orbit calculation. A well-depth
$V_{0}$ = 57.5 MeV is obtained for a
binding energy of 1.86 MeV relative to
the $J^{\pi} = 1/2^{+}$ state in ${}^{13}$C
at 3.09 MeV.

These parameters are slightly different 
from the parameters used in the reference \cite{Mex95}. 
The resulting DRC cross section calculations 
obtained using this set of parameters differ 
from the previous set of calculations by less 
than 5\%. This set has been adopted
in order to compare the DRC calculation with the
distorted waves Born approximation (DWBA)
employed in the analysis of the $(d,p)$ reaction, 
used to determine the spectroscopic factors
of the bound states of ${}^{13}$C \cite{Ohx85}.

The measured cross sections at $E_{n} = $ 550 keV
agree well with the calculation
based on the non-resonant DRC mechanism,
as was already shown in the low energy region 
analysis \cite{Mex95}. The results are shown
in Table ~\ref{tab2} and Fig.~\ref{fig7}. 
Here we confirm the interpretation of the DRC process 
and we extend it to higher energy.

From the measured cross section 
leading to the $J^{\pi} = 1/2^{+}$ state we
can derive the spectroscopic factor for
the $\mid {}^{12}$C$(0^{+}) \otimes (\nu 2s_{1/2})_{1/2^{+}} >$
configuration of this state. 
The spectroscopic factor turns out to be $S_{b}$ = 0.80, 
to be compared with the value derived from the
$(d,p)$ transfer reaction, $S_{b}$ = 0.65 \cite{Ohx85}.

In general, a 20\% uncertainty in the spectroscopic
factor determined from $(d,p)$ transfer reaction is
not uncommon. This uncertainty comes from various
factors, including the optical model-dependence 
of DWBA calculations, usually adopted 
in transfer reaction analysis. 
In our case, this uncertainty is
removed leaving the geometrical parameters of the
bound-state Woods-Saxon potential as the only
model-dependent uncertainty. This 
model dependency cannot be removed 
as it is inherent to the definition 
of spectroscopic factor of a
single-particle configuration, such as the  
$\mid {}^{12}$C$(0^{+}) \otimes (\nu 2s_{1/2})_{1/2^{+}} >$,
involved in the present case.

Assuming the experimental uncertainty of the 
measured capture cross section at 550 keV (5\%) 
as the only uncertainty, our analysis
results in $S_{b}$ = 0.80 $\pm$ 0.04 for the 
$J^{\pi} = 1/2^{+}$ state in ${}^{13}$C.

The same procedure could be applied to the
other transition, the one leading to the
$J^{\pi} = 5/2^{+}$ state of ${}^{13}$C
at 3.85 MeV. However, the measured capture cross
section leading to this state suffers from large
uncertainty and a meaningful determination of
the spectroscopic factor for this transition
cannot be performed at present.

In the case of the negative-parity states
(ground and second excited states of ${}^{13}$C), 
the uncertainty deriving from the assumed
model potential used to calculate the
collision matrix of the scattering channel
does not allow for a determination of the
spectroscopic factor with a better accuracy
than that obtained by the $(d,p)$ analysis.
This can be easily seen in Fig.~\ref{fig7}
where a calculation performed using a
simple plane-wave in the incident channel
results in a capture cross section
two order of magnitude larger than
the experimental values.

In our calculations we have adopted
the values of $S_{b}$ derived from the
$(d,p)$ reaction \cite{Ohx85} for the two
negative-parity states in ${}^{13}$C. Still, we would
like to stress here that using the same potential
parameters obtained to reproduce the bound state at
3.09 MeV in calculating the collision matrix
for {\it s}- and {\it d}-waves, we have obtained 
a very good agreement with the measured 
capture cross section in the whole energy range.
The calculated thermal ($E_{n}$ = 0.0253 eV)
capture cross section, due to these two transitions 
turns out to be
$\sigma_{n,\gamma}^{th}$ = 3.15 mb, to be compared
with the experimental 
$\sigma_{n,\gamma}^{th} = 3.53 \pm 0.07 $ mb \cite{Mugx81}.
Our interaction is also able to reproduce the
thermal scattering radius $R'$ = 6.45 fm,
to be compared with the experimental $R' = 6.3 \pm 0.1 $ fm \cite{Mugx81}.
These results can be considered satisfying as
the sensitiveness of the matrix elements to 
the potential employed to calculate the
collision matrix $U_{l=0}$ and the
uncertainty in the spectroscopic factors 
do not generally allow for better accuracy.

From the data shown in Fig.~\ref{fig7}
one can clearly see the onset of the capture
of {\it d}-wave neutrons. In fact, for incident
energies up to $\approx$ 100 keV, the {\it s}-wave
component of the incident channel wave function
is clearly dominant. For higher energies and in
particular around 500 keV, the contribution
of this higher-order partial wave becomes dominant.
To date, this is the first observation of the
onset of DRC {\it d}-wave neutron capture.

Finally, it should be added that, in general 
the presence of resonances near the threshold 
can make it difficult to compare the measured 
cross sections with the DRC calculation,
due to the possible interference effects between 
the resonant and the DRC processes.   
In the ${}^{12}$C$(n,\gamma)^{13}$C reaction, however, 
the nearest resonance state is rather high, 
at 2.08 MeV ($E_{n}$ = 2.25 MeV) 
with a narrow $\Gamma  = 8 \pm 2 $ keV width \cite{Mugx81}.
Thus, we could deduce the reaction mechanism 
and nuclear structure information 
more in detail using neutrons of $E_{n}$ between 
100 keV to 1 MeV, without worrying about 
possible interference effects. 

\subsection{$(n,\gamma)$ reaction rate calculation}
Using the present results just described
we can deduce the  ${}^{12}$C$(n,\gamma)^{13}$C 
reaction rate. This reaction rate is important for
several applications in nuclear astrophysics,
including the studies of inhomogeneous big-bang
models\cite{Ka95}, the nucleosynthesis of heavy elements
through the s-process\cite{Rax93} and r-process\cite{Hox93}. 
The Maxwellian averaged neutron capture cross section
for a temperature $T$ ($k$ is the Boltzmann constant), 
is given by
\begin{equation}
\frac{\langle \sigma v\rangle_{kT}}{v_T}
= \frac{2}{\sqrt{\pi}} \frac{1}{(kT)^2}
\int\limits_{0}^{\infty} E \ \sigma_{n,\gamma}(E)
\ {e}^{-E/kT} dE 
\end{equation}
where $v_{T} = (2kT/\mu)^{1/2}$ is the
velocity corresponding to the thermal energy $kT$. 
A short-hand notation for this quantity
often adopted is
$\langle \sigma \rangle_{kT}$. Indicating
with $\langle \sigma \rangle_{kT}^{s}$,
$\langle \sigma \rangle_{kT}^{p}$ and
$\langle \sigma \rangle_{kT}^{d}$
the respective components of the total capture
cross section due to incident {\it s}-,
{\it p}- and {\it d}-waves, we obtain
\begin{equation}
\langle \sigma \rangle_{kT}^{s} = \frac{17.3}{\sqrt{kT}}, \qquad
\langle \sigma \rangle_{kT}^{p} = 2.01 \sqrt{kT} 
\qquad \mbox{and} \qquad
\langle \sigma \rangle_{kT}^{d} = 6.0 \times 10^{-4} (kT)^{3/2}
\end{equation}
where the cross sections are in units of $\mu$b when $kT$
is in units of keV.
The {\it s}-wave component has been multiplied by a factor
of 1.12 to take into account the discrepancy between the
calculated and experimental thermal cross section (see above).

The resulting reaction rate is
\begin{eqnarray}
N_{A} \langle \sigma v \rangle & = & 
2.753 \times 10^{7} \  
\sqrt{kT} \ \langle \sigma \rangle_{kT} \nonumber \\ 
& = & 476.0  \quad + \quad
4765.2 \ T_{9} \quad + \quad 122.5 \ T_{9}^{2}  \qquad
\mbox{cm}^{3} \ \mbox{mole}^{-1} \ \mbox{sec}^{-1}.
\end{eqnarray}
The usual notation, $T_{9}$, for the temperature in units
of $10^{9} \ {}^{\circ}$K and  $N_{A}$ for the Avogadro 
constant has been adopted.
To show the relevance of the present result in comparison
with the previously adopted rate we show 
in Fig.~\ref{fig8} a comparison of the reaction rates 
responsible for the ${}^{12}$C processing for 
temperatures ranging from $10^{7}$ to $10^{10} \ {}^{\circ}$K.
All the reaction rates, except for our
present calculation, have been obtained from the
Nuclear Astrophysics Data service\cite{NADweb},
mainly adopting the rates of Caughlan and Fowler\cite{CF88}.
It can be seen that the inclusion of the {\it p}-wave
component in the capture cross section, resulting
in a linear increase of the reaction rate with
temperature, makes the neutron capture
competing and in most cases exceeding the other
processing mechanisms for temperatures of
the full range. The consequences of this result
on stellar as well as on the big-bang
nucleosynthesis modeling is presently 
under investigation.

\section{Conclusion}
The present work demonstrates clearly the important 
r\^{o}le of prompt discrete $\gamma$-ray detection technique
using a high resolution NaI(Tl) spectrometer
for understanding nuclear structure properties 
and capture reaction mechanisms.
We succeeded for the first time 
to measure the partial cross sections from the
${}^{12}$C$(n,\gamma)^{13}$C reaction at $E_{n}$ = 550 keV.  
By comparing the measured values with a calculation
based on a non-resonant DRC model with various
partial wave components, 
we have been able to determine the spectroscopic factor 
for the $J^{\pi} = 1/2^{+}$ 3.09 MeV state in ${}^{13}$C,
reliably. In addition we have
unambiguously identified the onset of 
the {\it d}-wave neutron capture for incident
neutron energies above 100 keV.
The reaction rate for the 
${}^{12}$C$(n,\gamma)^{13}$C reaction obtained by our
work differs substantially from the values adopted so
far, producing interesting consequences on the
astrophysical applications of our results.

\vspace{2.0pc}
We acknowledge fruitful discussions with 
M.~Ishihara, H.~Ohnuma and M.~Hashimoto. 
This work was supported by a grant-in-aid for Specially Promoted
Research of the Japan Ministry of Education, Science, Sports
and Culture and partly by a grant-in-aid 
for Scientific Research on Priority Areas.

\vspace{2.0pc}



\clearpage

\clearpage

\clearpage

\clearpage
\begin{figure}
\begin{center}
\leavevmode
\hbox{
\epsfxsize=13.0cm
\epsfysize=8.7cm
\epsffile{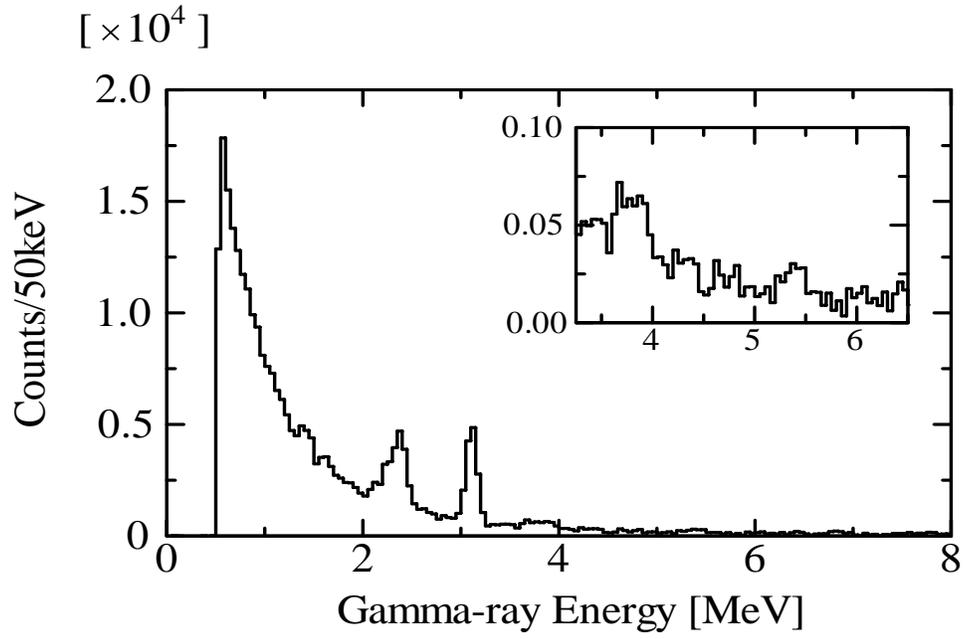}}
\end{center}
\caption{Background subtracted $\gamma$-ray spectra 
for the  ${}^{12}$C$(n,\gamma)^{13}$C reaction.}
\label{fig4}
\end{figure}

\clearpage
\begin{figure}
\begin{center}
\leavevmode
\hbox{
\epsfxsize=13.0cm
\epsfysize=10.0cm
\epsffile{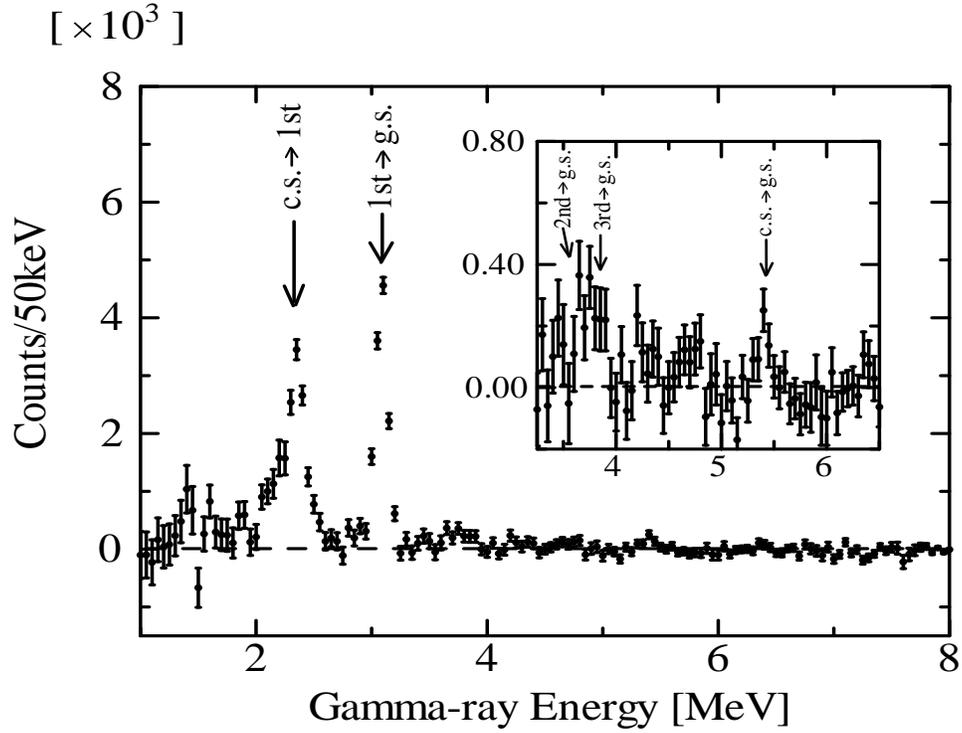}}
\end{center}
\caption{Net $\gamma$-ray spectrum 
for the ${}^{12}$C$(n,\gamma)^{13}$C reaction. 
This spectrum was obtained by subtracting the $\gamma$-ray spectrum
measured without the C sample from the background subtracted
spectrum shown in \protect\ref{fig4}.}
\label{fig5}
\end{figure}

\clearpage
\begin{figure}
\begin{center}
\leavevmode
\hbox{
\epsfxsize=13.0cm
\epsfysize=8.7cm
\epsffile{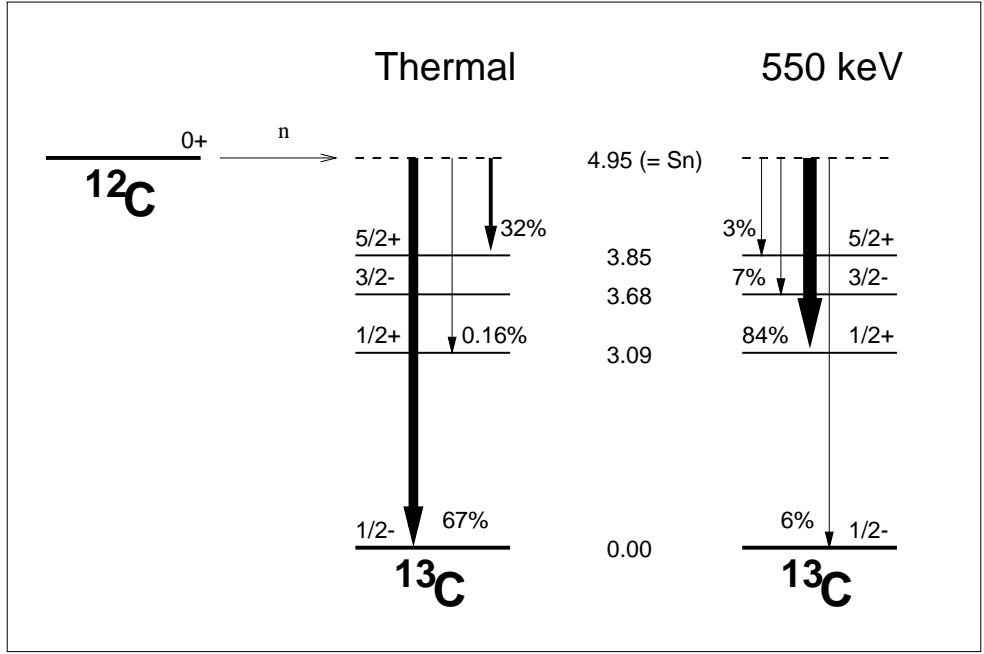}}
\end{center}
\caption{Partial level scheme of ${}^{13}$C. The capture
$\gamma$-ray branching ratios for incident thermal
neutrons (left) and for the fast, 550 keV neutrons (right)
are shown together with the excitation energies (in MeV).}
\label{fig6}
\end{figure}

\clearpage
\begin{figure}
\begin{center}
\leavevmode
\hbox{
\epsfxsize=13.0cm
\epsfysize=11.7cm
\epsffile{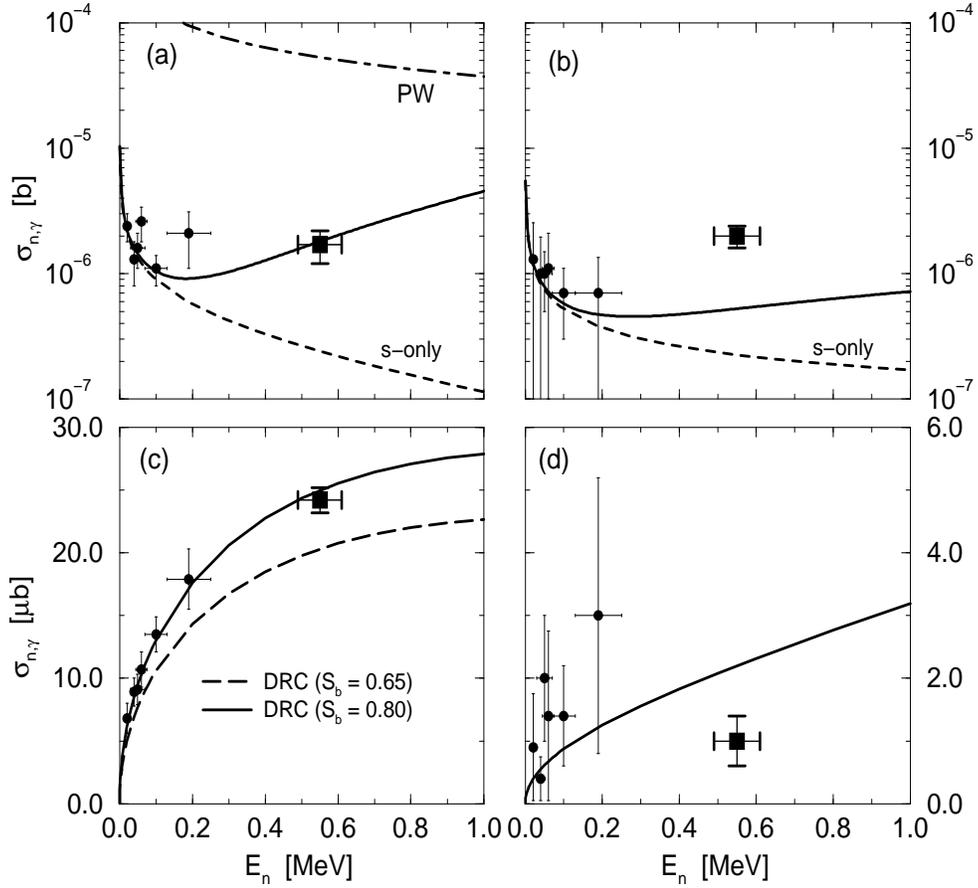}}
\end{center}
\caption{Neutron capture cross section of ${}^{12}$C for
(a) transitions leading to the ${}^{13}$C ground-state,
(b) to the second-excited, (c) to the first-excited  
state and (d) to the third-excited state. 
In (a) and (b)
the dashed line indicates the incident {\it s}-wave
contribution only and the full line the total
({\it s}+{\it d}-wave) DRC. In (a), the line labelled with
PW indicates a calculation made with a plane-wave
wave function (see \protect\cite{Mex95} for details).
In (c), the dashed line indicates
the result obtained using a spectroscopic factor
$S_{b}$ = 0.65 whereas the full line indicates the
result obtained using $S_{b}$ = 0.80. The experimental
values for $E_{n}$ = 550 keV (square symbols) 
are from the present
experiment while the other experimental values
are from a previous experiment \protect\cite{Ohx94}.}
\label{fig7}
\end{figure}

\clearpage
\begin{figure}
\begin{center}
\leavevmode
\hbox{
\epsfxsize=13.0cm
\epsfysize=11.7cm
\epsffile{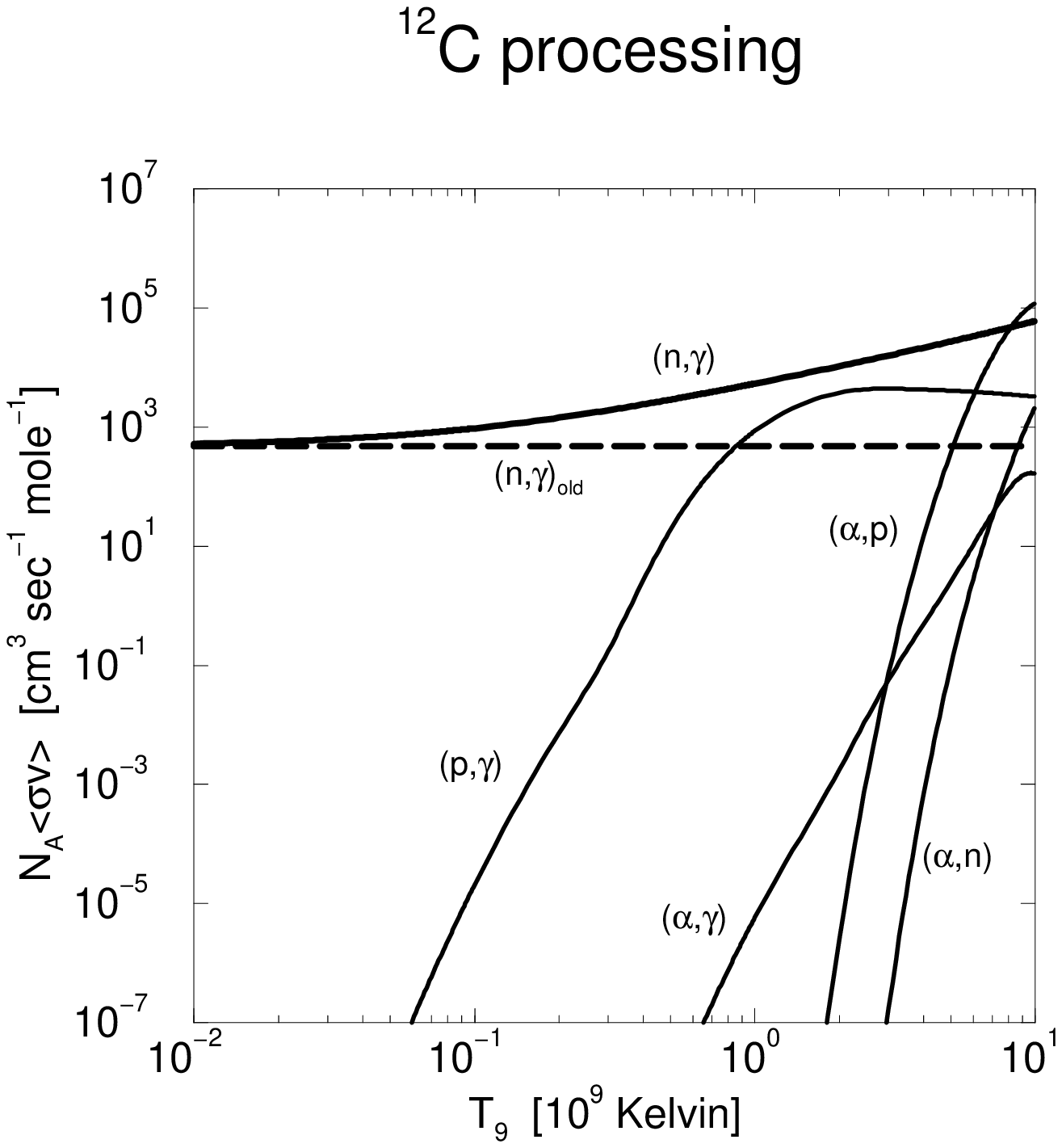}}
\end{center}
\caption{The reaction rates for processing ${}^{12}$C 
from various reaction channels for temperatures ranging from
$10^{7}$ to $10^{10} \ {}^{\circ}$K 
($T_{9} \equiv 10^{9} \ {}^{\circ}$K).
Except for the $(n,\gamma)$ rate, the rates are from the
Nuclear Astrophysics Data Service \protect\cite{NADweb}.
The dashed line, labelled by $(n,\gamma)_{old}$,
corresponds to a neutron capture cross 
section $\sigma_{n,\gamma}^{th} = 3.53$ mb 
and based on {\it s}-wave capture only. The increase
with temperature of the neutron capture rate
is due to the {\it p}- and {\it d}-wave DRC process.
See text for details.}
\label{fig8}
\end{figure}


\clearpage
\begin{table}
\caption{Correction factors for {\it s}- and {\it p}-wave neutrons.}
\begin{tabular}{c l c c c} 
Sample &          & $C_{n m}$ & $C_{n s}$ & $C_{n m} \times C_{n s}$ \\
\tableline
 C & {\it s}-wave &  1.87  &  0.68  &  1.27 \\
   & {\it p}-wave &  1.65  &  0.68  &  1.13 \\
Au & {\it s}-wave &  1.07  &  0.98  &  1.05 \\
\end{tabular}
\label{tab1}
\end{table}

\clearpage
\begin{table}
\caption{Measured and calculated capture cross sections ($\mu$barn)
at $E_{n}^{lab}$ = 550 keV.}
\begin{tabular}{l l l l} 
Bound-state $E_{x}$ (MeV)              & 
$E_{\gamma}$ (MeV)                     & 
$\sigma_{n,\gamma}^{exp}$ ($\mu$barn)  &
$\sigma_{n,\gamma}^{DRC}$ ($\mu$barn) \\
\tableline
0.0  &  5.45 &  1.7 $\pm$ 0.5 &  1.8 \\
3.09 &  2.36 & 24.2 $\pm$ 1.0 & 25.0\tablenote{Obtained
using $S_{b}$ = 0.80.} \\
3.68 &  1.77 &  2.0 $\pm$ 0.4 &  0.5 \\
3.85 &  1.60 &  1.0 $\pm$ 0.4 &  2.2 \\
\end{tabular}
\label{tab2}
\end{table}

\clearpage

\section*{FIGURE CAPTIONS}

\begin{itemize}
\item FIGURE 1: Neutron energy spectrum measured by a
${}^{6}$Li-glass scintillation counter with a TOF method.
The spectrum has been corrected for the energy-dependendent
neutron detection efficiency of the ${}^{6}$Li-glass.

\item FIGURE 2: Time-of-flight spectrum measured by the NaI(Tl)
detector for the Au sample. The sharp peak at channel 440 is
due to the $\gamma$-ray from the ${}^{7}$Li$(p,\gamma)^{8}$Be
reaction. The regions F and B were chosen to obtain the foreground 
and background $\gamma$-ray spectra, respectively (see text).

\item FIGURE 3: Foreground (solid line) and background (dotted line)
$\gamma$-ray spectra for the  ${}^{12}$C$(n,\gamma)^{13}$C. 
Iron was used to cover the NaI(Tl) spectrometer.

\item FIGURE 4: Background subtracted $\gamma$-ray spectra 
for the  ${}^{12}$C$(n,\gamma)^{13}$C reaction.

\item FIGURE 5: Net $\gamma$-ray spectrum 
for the ${}^{12}$C$(n,\gamma)^{13}$C reaction. 
This spectrum was obtained by subtracting the $\gamma$-ray spectrum
measured without the C sample from the background subtracted
spectrum shown in \ref{fig4}.

\item FIGURE 6: Partial level scheme of ${}^{13}$C. The capture
$\gamma$-ray branching ratios for incident thermal
neutrons (left) and for the fast, 550 keV neutrons (right)
are shown together with the excitation energies (in MeV).

\item FIGURE 7: Neutron capture cross section of ${}^{12}$C for
(a) transitions leading to the ${}^{13}$C ground-state,
(b) to the second-excited, (c) to the first-excited  
state and (d) to the third-excited state. 
In (a) and (b)
the dashed line indicates the incident {\it s}-wave
contribution only and the full line the total
({\it s}+{\it d}-wave) DRC. In (a), the line labelled with
PW indicates a calculation made with a plane-wave
wave function (see \cite{Mex95} for details).
In (c), the dashed line indicates
the result obtained using a spectroscopic factor
$S_{b}$ = 0.65 whereas the full line indicates the
result obtained using $S_{b}$ = 0.80. The experimental
values for $E_{n}$ = 550 keV (square symbols) 
are from the present
experiment while the other experimental values
are from a previous experiment \cite{Ohx94}.

\item FIGURE 8: The reaction rates for processing ${}^{12}$C 
from various reaction channels for temperatures ranging from
$10^{7}$ to $10^{10} \ {}^{\circ}$K 
($T_{9} \equiv 10^{9} \ {}^{\circ}$K).
Except for the $(n,\gamma)$ rate, the rates are from the
Nuclear Astrophysics Data Service \cite{NADweb}.
The dashed line, labelled by $(n,\gamma)_{old}$,
corresponds to a neutron capture cross 
section $\sigma_{n,\gamma}^{th} = 3.53$ mb 
and based on {\it s}-wave capture only. The increase
with temperature of the neutron capture rate
is due to the {\it p}- and {\it d}-wave DRC process.
See text for details.
\end{itemize}

\end{document}